\documentclass[twoside,12pt]{article}
\usepackage{epsfig}

\def\Journal#1#2#3#4{{#1} {#2} (#4) #3 }

\def\NPA{{\em Nucl. Phys.} A}

\def\PRL{\em Phys. Rev. Lett.}

\def\PRC{{\em Phys. Rev.} C}

\newcommand{\be}{\begin{equation}}
\newcommand{\ee}{\end{equation}}
\newcommand{\bea}{\begin{eqnarray}}
\newcommand{\eea}{\end{eqnarray}}

\begin{document}

\title{ \vspace{1cm} Constraints on the density dependence of the symmetry energy from heavy ion collisions}
\author{M.B.\ Tsang,$^1$ Z.\ Chajecki,$^1$ D.\ Coupland,$^1$ P.\
Danielewicz,$^{1}$ F.\ Famiano,$^2$ R.\ Hodges,$^1$\\
M.\ Kilburn,$^1$ F.\ Lu,$^3$ W.G. \ Lynch,$^1$ J.\
Winkelbauer,$^1$
M.\ Youngs,$^1$ Y.X. \ Zhang$^4$\\
\\
$^1$ National Superconducting Cyclotron Laboratory \\
Michigan State University, East Lansing, MI 48824, USA\\
$^2$ Physics Dept, Western Michigan University, Kalamazoo, MI USA\\
$^3$ School of Physics, Peking University, Beijing 100871, China\\
$^4$ China Institute of Atomic Energy, Beijing 102413, China\\}

\maketitle
\begin{abstract} Constraints on the Equation of State for symmetric matter (equal neutron and proton numbers) have been extracted from energetic collisions of heavy ions over a range of energies. Collisions of neutron-deficient and neutron-rich heavy ions now provide initial constraints on the EoS of neutron-rich matter at sub-saturation densities from isospin diffusions and neutron proton ratios. This article reviews the experimental constraints on the density dependence of Symmetry Energy at sub-saturation density.
\end{abstract}
\section{Introduction}
The Equation of State (EoS) of cold nuclear matter can be written
as the sum of the energy per nucleon of symmetric matter and a
symmetry energy term, $E_\delta$
\begin{equation}
E(\rho,\delta)=E_0 (\rho, \delta=0)+E_\delta
\end{equation}
where $\delta=(\rho_n-\rho_p)/\rho$ is the asymmetry; $\rho_n$,
$\rho_p$ and $\rho$ are the neutron, proton and nucleon densities,
respectively. The first term on the RHS, $E_0(\rho,\delta=0)$, is
the EoS term for symmetric nuclear matter with equal fractions of
neutrons and protons. Significant constraints on the symmetric
matter EoS at 1 $\le\rho/\rho_0\le$ 4.5 have been obtained from
measurements of collective flow \cite{Dan02} and Kaon production
\cite{Fuc06}. However, well-determined constraints on the symmetry
term, $E_\delta=S(\rho,\delta^2)$, are few. $S(\rho)$ describes
the density dependence of $E_\delta$.

Since large variations in nuclear density can be attained
momentarily in nuclear collisions, constraints on the Equation of
State (EoS) can be obtained by comparing measurements to transport
calculations of such collisions. The symmetry energy has been
recently probed at sub-saturation densities via isospin diffusion
\cite{Tsa04, Liu07}, and by double ratios involving neutron and
proton energy spectra \cite{Fam06}. These two observables largely
reflect the transport of nucleons under the combined influence of
the mean fields and the collisions induced by residual
interactions; thus, they should be within the predictive
capabilities of transport theory.

This article focuses on investigations of the symmetry energy at
sub-saturation density. The constraints obtained in heavy ion
collisions are compared to constraints obtained from nuclear
structure studies, from Isobaric Analog States, Sn isotope skins,
Pygmy Dipole Resonance experiments, and the Giant Dipole Resonance
data. I will briefly discuss investigations of the density
dependence of symmetry energy at supranormal density.

\section{ImQMD: Transport calculations for heavy ion collisions}

Transport models have been used to describe the dynamics of heavy
ion collisions. We chose to use the ImQMD code mainly because of
its ability to describe fragments which are quite important in
describing isospin observables. In  the QMD model, nucleons are
represented by Gaussian wavepackets. The mean fields acting on
these wavepackets are derived from an energy functional with the
potential energy U that includes the full Skyrme potential energy
with only the spin-orbit term omitted:
\begin{equation}
U=U_{\rho}+U_{md}+U_{coul} \label{upot};U_{\rho,md}=\int
u_{\rho,md} d^{3}r \label{urhomd}
\end{equation}
where, $U_{coul}$ is the Coulomb energy and $U_{\rho,md}$
represents the nuclear contributions
 in local form and
\begin{equation}
u_{\rho}=\frac{\alpha}{2}\frac{\rho^{2}}{\rho_{0}}+\frac{\beta}{\eta+1}\frac{\rho^{\eta+1}}{\rho^{\eta}_{0}}+\frac{g_{sur}}{2\rho_{0}}(\nabla
\rho)^2 \nonumber+\frac{g_{sur,iso}}{\rho_{0}}(\nabla
(\rho_{n}-\rho_{p}))^2
\nonumber+\frac{C_{s}}{2}(\frac{\rho}{\rho_{0}})^{\gamma_i}\delta^{2}\rho+g_{\rho\tau}\frac{\rho^{8/3}}{\rho_{0}^{5/3}}
\label{urho}
\end{equation}
where the asymmetry $\delta$, and $\rho_{n}$ and $\rho_{p}$ are the neutron and proton densities, respectively. A symmetry kinetic energy density of the form $\frac{C_{s,k}}{2}(\frac{\rho}{\rho_0})^{2/3}\delta^2\rho$ and symmetry potential energy density of the form $\frac{C_{s,p}}{2}(\frac{\rho}{\rho_0})^{\gamma}\delta^2\rho $ were used in the following transport model comparisons. The energy density associated with the mean-field momentum dependence is represented by
\begin{equation}
u_{md}=\frac{1}{2\rho_{0}}\sum_{N_1,N_2} \frac{1}{16\pi^{6}}\int
d^{3}p_{1}d^{3}p_{2}f_{N_{1}}(\vec{p}_1)f_{N_{2}}(\vec{p}_2)\nonumber
1.57[\ln(1+5\times 10^{-4}(\Delta p)^2)]^2 \label{eq:umd}
\end{equation}
where $f_N$ are nucleon Wigner functions, $\Delta p=|\vec{p_1}-\vec{p_2}|$, the energy is in MeV and momenta are in MeV/c. The resulting interaction between wavepackets is described in Ref. \cite{Aic87}. Unless otherwise noted, we use $\alpha$=-356 MeV, $\beta$=303 MeV and $\eta$ =7/6, corresponding to a isoscalar compressibility constant of K=200 MeV, and $g_{sur}$=19.47 MeVfm$^2$, $g_{suriso}$=-11.35 MeVfm$^2$, $C_{s,k}$=24.9 MeV, $C_{s,p}$=35.19 MeV, and $g_{\rho\tau}$ =0 MeV.

These calculations use isospin-dependent in-medium nucleon-nucleon scattering cross sections in the collision term and Pauli blocking effects that are described in \cite{Zha06,Zha07}. Cluster yields are calculated by means of the coalescence model widely used in QMD calculations in which particles with relative momenta smaller than $P_{0}$ and relative distances smaller than $R_{0}$ are coalesced into one cluster. In the present work, values of $R_{0}=3.5fm$ and $P_{0}=250MeV/c$ are employed.

\section{Experimental constraints from heavy ion collisions}
We turn our attention first to the interpretation of neutron/proton double ratio data, which derives its sensitivity to the symmetry energy from the opposite sign of the symmetry force for neutrons as compared to protons \cite{Li97}. First experimental comparisons of neutron and proton spectra in Ref.\cite{Fam06} used a double ratio in order to reduce sensitivity to uncertainties in the neutron detection efficiencies and to relative uncertainties in energy calibrations of neutrons and protons. This double ratio,
\begin{equation}
DR(n/p) = R_{n/p}(A)/ R_{n/p}(B) = \frac{dM_n(A)/dE_{c.m.}}{dM_p(A)/dE_{c.m.}}\cdot\frac{dM_n(B)/dE_{c.m.}}{dM_p(B)/dE_{c.m.}},
\end{equation}
was constructed by measuring the energy spectra, $dM/dE_{c.m.}$, of neutrons and protons for two systems A and B characterized by different isospin asymmetries. The star symbols in the left panel of Figure 1 show the double ratios measured at $70^{o}\le\theta_{CM}\le110^{o}$ as a function of nucleon center-of-mass (c.m.) energy for central collisions of $^{124}Sn+^{124}Sn$ and $^{112}Sn+^{112}Sn$ at E/A=50 MeV.

We have performed calculations at an impact parameter of b=1, 2
and 3 fm at an incident energy of 50 MeV per nucleon for two
systems: A=$^{124}Sn+^{124}Sn$ and B=$^{112}Sn+^{112}Sn$. About
60,000 events are simulated for each impact parameter. Within the
statistical uncertainties, the double ratio observable, DR(n/p),
is nearly independent of the impact parameters within the range of
0$\le$b$\le$5 fm. The lines in the left panel of Figure 1 show the
predicted double ratios DR(np)=$R_{n/p}(124)/R_{n/p}(112)$ as a
function of the c.m. energy of nucleons emitted at
$70^{o}\le\theta_{C.M.}\le110^{o}$ for $\gamma_i$=0.35, 0.5, 0.75,
1 and 2. The uncertainties for these calculations in Figure 1 are
statistical. As fewer nucleons are emitted at high energy, the
uncertainties increase with increasing energy. Despite the large
experimental uncertainties, especially for the higher energy
($>$40 MeV) data, the trends and magnitudes of the data points
definitely rule out the very soft ( $\gamma_i$=0.35, dotted line
with open diamond points) and very stiff ( $\gamma_i$=2, dotted
line with closed diamond symbols) density-dependent symmetry
terms. We find the $2\sigma$ uncertainty range of values for
$\gamma_i$ to be 0.4$\le \gamma_i\le$ 1.04, corresponding to an
increase in $\chi^2$ by 4 above its minimum of $\chi^2 \sim$ 2.1
near $\gamma_i=0.7$.

\begin{figure}[tb]
\begin{center}
\begin{minipage}[t]{8 cm}
\epsfig{file=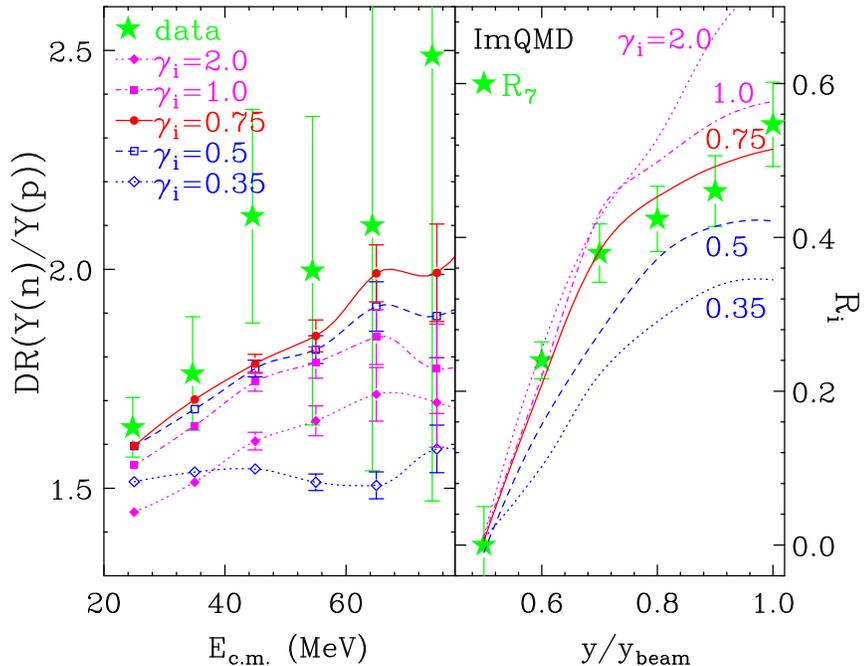,scale=0.5,angle=90}
\end{minipage}
\begin{minipage}[t]{16.5 cm}
\caption{(Color online): Left panel: Comparison of experimental
double neutron-proton ratios (star symbols), as a function of
nucleon center of mass energy, to ImQMD calculations (lines) with
different density dependence of the symmetry energy parameterized
in terms of $\gamma_i$ in Eq. (1). Right panel: Comparison of
experimental isospin transport ratios obtained from the yield of
A=7 isotopes (star symbols), as a funtion fo the beam rapidity, to
ratios from ImQMD calculations (lines) for b=6 fm.\label{fig1}}
\end{minipage}
\end{center}
\end{figure}

The density dependence of the symmetry energy can also be probed in peripheral collisions between two nuclei with different isospin asymmetries by examining the diffusion of neutrons and protons across the neck that joins the nuclei. This "isospin diffusion" generally continues until the two nuclei separate or until the chemical potentials for neutrons and protons in both nuclei become equal. To isolate diffusion effects from other effects such as pre-equilibrium emission, Coulomb effects, and secondary decays, measurements of isospin diffusion compare "mixed" collisions involving a neutron-rich nucleus A and a neutron-deficient nucleus B to the "symmetric" collisions involving A+A and B+B. The degree of isospin equilibration in such collisions can be quantified by rescaling the isospin observable X according to the isospin transport ratio $R_i(X)$ \cite{Tsa04} given by
\begin{equation}
R_i(X)=2\frac{X-(X_{A+A}+X_{B+B})/2}{X_{A+A}-X_{B+B}}
\end{equation}
In the absence of isospin diffusion, $R_i(X_{A+B})=R_i(X_{A+A})$=1
for a collision involving a neutron-rich projectile on a
proton-rich target. Likewise, without diffusion,
$R_i(X_{B+B})=R_i(X_{B+A})$=-1. On the other hand, if isospin
equilibrium is achieved for roughly equal sized projectiles and
target nuclei, $R_i(X_{A+B})=R_i(X_{B+A})\approx 0$. Using ImQMD
model, we investigate $R_i(\delta)$ as a function of the rapidity
of the emitted particles. The results of the calculations are
shown by the lines in the right panel of Figure 2. The data,
represented by the star symbols, are obtained from the yield
ratios of the mirror nuclei pair, $^7Li$ and $^7Be$, $R_7=
R_i(X=ln(Y(^7Li)/Y(^7Be)))$\cite{Liu07}. This calculation
reproduces the shapes and magnitudes of the rapidity dependence of
the isospin transport ratios $R_7$. The $\chi^2$ analysis brackets
$\gamma_i$ values to lie in the region 0.4$\le\gamma_i\le$1.0
using the same $2\sigma$ criterion. There is substantial overlap
between the constraints obtained from isospin diffusion and double
neutron and proton yield ratios.

Constraints on the exponent $\gamma_i$ depend on the symmetry
energy at saturation density, $S_0=
S(\rho_0)=0.5*(C_{s,k}+C_{s,k})$. Increasing $S_0$ has the same
effect on the isospin transport ratio as decreasing $\gamma_i$.
Thus $S_0$ and $\gamma_i$ cannot be determined independently. To
provide constraints on $S_0$ and $\gamma_i$ parameters, contours
with constant $\chi^2$ values can be created by doing
two-dimensional $\chi^2$ analysis in the $\gamma_i - S_0$
parameter space. Unfortunately, the CPU requirements for such
large-scale ImQMD calculations are too intensive to make such
analysis practical. Instead, We have preformed a series of ImQMD
calculations at b=6 fm with different values of $\gamma_i$ over
selected values of $S_0$ between 25 to 40 MeV to locate the
approximate boundaries in the $S_0$ and $\gamma_i$  plane that
satisfy the 2$\sigma$ criterion in the $\chi^2$ analysis of the
isospin diffusion data. The shaded area shown in the inset of
Figure 3 is the result of such analysis.

\section{ Comparison of experimental constraints at sub-saturation density}
Near saturation density, one may expand the symmetry energy, $S(\rho)$, about the saturation density, $\rho_0$,
\begin{equation}
    S(\rho)=S_0+\frac{L}{3}\frac{\rho-\rho_0}{\rho_0}+\frac{K_{sym}}{18}(\frac{\rho-\rho_0}{\rho_0})^2+ \cdot \cdot \cdot
\end{equation}
where L and $K_{sym}$ are slope and curvature parameters at $\rho_0$.  The slope parameter, L, is related to $p_0$, the pressure from the symmetry energy for pure neutron matter at saturation density:
\begin{equation}
    L=3\rho_0 |dS(\rho)/d\rho|_{\rho_0}= [3/\rho_0]\cdot p_0
\end{equation}
$p_0$, provides the baryonic contribution to the pressure in neutron stars \cite{Ste05}, where the energy of symmetric matter, $E_0(\rho,\delta=0)$, does not contribute to the pressure. It is also related to the neutron skin thickness ( $\delta R_{np}$) of neutron-rich heavy nuclei including $^{208}Pb$ \cite{Hor01,Typ01}.

The series of ImQMD results calculated at b=6 fm with different
values of $\gamma_i$ over selected values of $S_0$ between 25 to
40 MeV give the approximate boundaries in the $S_0$ and $L$ plane
that satisfy the 2$\sigma$ criterion in the $\chi^2$ analysis of
the isospin diffusion data. The area bounded by the two solid
diagonal lines in Figure 2 and the shaded region in Figure 3
represents a conservative estimate in such effort. Another
independent constraint will be needed to provide constraints in
$S_0$ or $L$.

\begin{figure}[tb]
\begin{center}
\begin{minipage}[t]{8 cm}
\epsfig{file=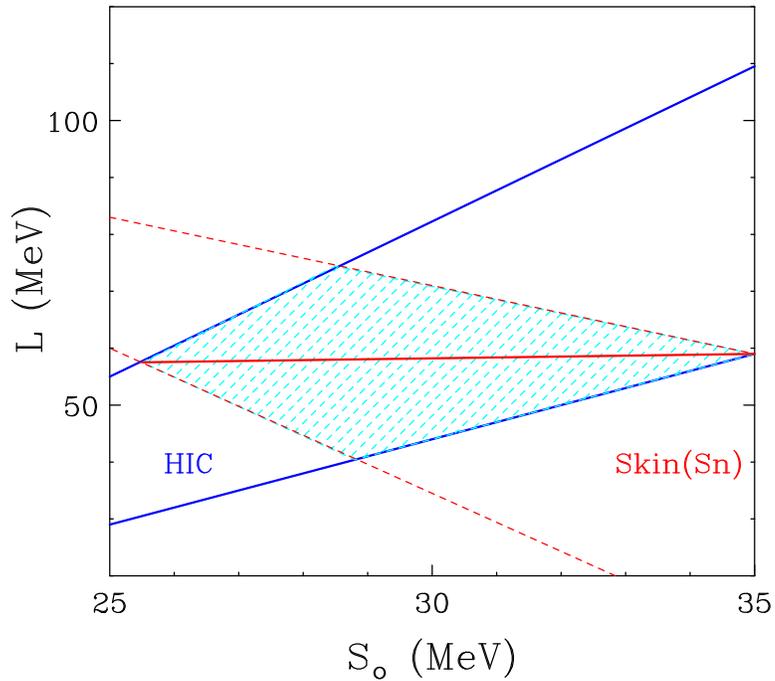,scale=0.5,angle=90}
\end{minipage}
\begin{minipage}[t]{16.5 cm}
\caption{  (Color online): Representation of the constraints on
the density dependence of symmetry energy as a function of the
symmetry energy at saturation density in neutron matter. The
hatched areas are extracted from recent analysis of nuclear
collisions of Sn nuclei. The two dashed lines represent the
constraint boundaries obtained from the analysis of the skin
thickness of tin isotopes \cite{Che10}. The horizontal line
indicates the nearly constant mean L value in the overlap
region.\label{fig3}}
\end{minipage}
\end{center}
\end{figure}

\begin{figure}[tb]
\begin{center}
\begin{minipage}[t]{8 cm}
\epsfig{file=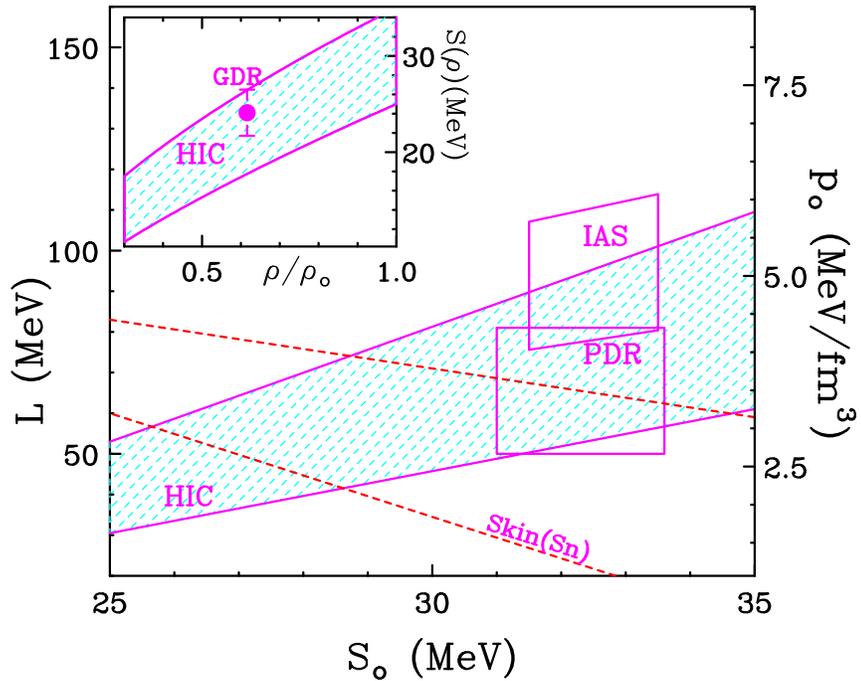,scale=0.5,angle=90}
\end{minipage}
\begin{minipage}[t]{16.5 cm}
\caption{ (Color online): Comparison of various constraints on the
density dependence of symmetry energy (inset) and pressure as a
function of the symmetry energy at saturation density in neutron
matter. The hatched areas are extracted from recent analyses of
nuclear collisions of Sn nuclei. Open rectangles and the symbol in
the inset were obtained from the properties of individual nuclei.
The inset shows the density dependence of the symmetry energy of
the shaded region. The symbol in the inset represents the GDR
results \cite{Tri08}.\label{fig4}}
\end{minipage}
\end{center}
\end{figure}

Recent analysis of the skin thickness of tin isotopes \cite{Che10}
suggests that L decreases with increasing $S_0$. The two diagonal
dashed lines plotted in Figure 2 represent a 2$\sigma$ analysis of
the $\chi^2$ discussed in Ref. \cite{Che10}. Since the trend is
opposite to that of the constraints provided by the heavy ion
reactions (solid lines), the overlap of the two constraints
provide a range of $S_0$ and $L$ values. As pointed out by Chen et
al, the interesting thing to note is that the mean $L$ values,
represented by the horizontal line in Figure 2 obtained from the
overlap region of the two constraints, are nearly constant.  The
range of $L$ and $S_0$ values in the overlap region corresponds to
a mean skin thickness of $^{208}Pb$ of about 0.18$\pm$0.04 fm
\cite{Ste05}. The range of experimental neutron skin thicknesses
of $^{208}Pb$ varies from about 0.13 to 0.22 fm \cite{Kra94,
Ray79, Kar02, Klo07, Sta94} which are consistent with results
predicted by the constraints.

We have included, in Figure 3, other constraints from the
properties of individual nuclei. The corresponding values of pure
neutron matter pressure, $p_0$, are given in the right axis. The
lower box centered at $S_0$ = 32.5 MeV depicts the range of $S_0$
and $\rho_0$ values from analysis of low-lying electric dipole
strength (Pygmy Dipole Resonance, PDR) in neutron-rich nuclei
$^{68}Ni$ and $^{132}Sn$ \cite{Kli07, Car10}. The upper box
centered at $S_0$ = 32.5 MeV depicts the constraints reported in
Ref. \cite{Dan09} from the analysis of the excitation energies of
Isobaric Analog states (IAS). All these constraints overlap
significantly with the hatched regions obtained from the heavy ion
collisions of tin isotopes as described in section 3. The inset
shows what this analysis implies about the density dependence of
the symmetry energy. A value for the symmetry energy at around 0.6
normal nuclear matter density, extracted from Giant Dipole
Resonance (GDR) data \cite{Tri08}, is indicated by the symbol in
the inset.

\section{Summary and outlook}

In summary, the ability of QMD models to reproduce both isospin
diffusion data and double ratio of neutron and proton spectra data
is an important step forward in obtaining information about the
symmetry energy in heavy-ion collisions. The heavy ion as well as
nuclear structure observables examined here provide consistent
constraints on the density dependence of the symmetry energy.

To explore the density dependence of the symmetry energy, an
international collaboration, consisting of an interdisciplinary,
experimental and theoretical team of scientists, has been formed
\cite{SEP10} to conduct a series of experiments at unique
facilities based in the United States (the National
Superconducting Cyclotron Laboratory at Michigan State
University), Japan (the Radioactive Ion Beam Factory at RIKEN) and
GSI, Germany. Each facility enables the exploration of a different
density range.  With experiments underway at GSI, better
constraints at supra normal density is expected in the near
future.

\textbf{Acknowledgements}

This work has been supported by the U.S. National Science
Foundation under Grants PHY-0606007, 0800026, and the Chinese
National Science Foundation of China under Grants 10675172.


\begin{thebibliography}{99}
\itemsep -2pt

\bibitem{Dan02} P. Danielewicz et al.,\Journal{\em Science} {298}{1592} {2002}
\bibitem{Fuc06} C. Fuchs, \Journal{\em Prog. Part. Nucl. Phys.} {56}{1} {2006}
\bibitem{Tsa04} M. B. Tsang et al., \Journal{\PRL} {92}{062701} {2004}
\bibitem{Liu07} T. X. Liu et al., \Journal{\PRC} {76}{034603} {2007}
\bibitem{Fam06} M. A. Famiano et al., \Journal{\PRL} {97} {052701} {2006}.
\bibitem{Aic87} J. Aichelin et al., \Journal{\PRL} {58} {1926} {1987}.
\bibitem{Zha06} Y. Zhang and Z. Li, \Journal{\PRC} {74} {014602} {2006}.
\bibitem{Zha07} Y. Zhang, Z. Li and P. Danielewicz, \Journal{\PRC} {75} {034615} {2007}.
\bibitem{Li97} B. A. Li et al., \Journal{\PRL} {78} {1644} {1997}.
\bibitem{Ste05} A. W. Steiner, \Journal{\em Phys. Rep.} {411}{325} {2005}
\bibitem{Hor01} C. J. Horowitz et al., \Journal{\PRL} {86} {5647} {2001}.
\bibitem{Typ01} S. Typel and B. A. Brown, \Journal{\PRC} {64} {027302} {2001}.
\bibitem{Che10} L.W. Chen et al., \Journal{\PRC} {82} {024321} {2010}
\bibitem{Kra94} A. Krasznahorkay et al., \Journal{\NPA} {567} {521} {1994}.
\bibitem{Ray79} L. Ray, \Journal{\PRC} {19} {1855} {1979}.
\bibitem{Kar02} S. Karataglidis et al., \Journal{\PRC} {65} {044306} {2002}.
\bibitem{Klo07} B. Klos et al., \Journal{\PRC} {76} {014311} {2007}.
\bibitem{Sta94} V. E. Starodubsky and N. M. Hintz, \Journal{\PRC} {49} {2118} {1994}.
\bibitem{Kli07} A. Klimkiewicz et al., \Journal{\PRC} {76}{051603} {2007}
\bibitem{Car10} Andrea Carbone et. al, \Journal{\PRC} {81} {041301(R)} {2010}.
\bibitem{Dan09} P. Danielewicz, Jenny Lee \Journal{\em Nucl. Phys. A} {818}{36} {2009}
\bibitem{Tri08} Luca Trippa et al., \Journal{\PRC} {77} {061304(R)} {2008}.
\bibitem{SEP10} http://groups.nscl.msu.edu/hira/sep.htm.

\end{thebibliography}
\end{document}